\documentclass[useAMS]{mn2e}
\usepackage{times,psfig,epsfig}
\usepackage{graphicx}
\usepackage{times}
\begin{document}
\date{Accepted 2006 August 31. Received 2006 Augsut 31; in original form 2006 April 10}
\title[Spectral indices of correlated astrophysical
foregrounds]{Estimating the spectral indices of correlated
astrophysical foregrounds by a second-order statistical approach}

\author[A. Bonaldi, L. Bedini, E. Salerno, C. Baccigalupi, G. De Zotti]
{A. Bonaldi$^{1,2}$, L. Bedini$^{3}$, E. Salerno$^{3}$, 
C.Baccigalupi$^{4,5}$, G. De Zotti$^{2,4}$ 
\\
$^{1}$Dipartimento di Astronomia, Universit\`a di Padova, vicolo dell'Osservatorio 2, I-35122 Padova, Italy \\
$^{2}$INAF-Osservatorio Astronomico di Padova, vicolo dell'Osservatorio 5, I-35122, Padova, Italy\\
$^{3}$Istituto di Scienza e Tecnologie dell'Informazione, CNR, Area della ricerca di Pisa, via G. Moruzzi 1, I-56124 Pisa, Italy\\
$^{4}$SISSA/ISAS, via Beirut 4, I-34014 Trieste, Italy\\
$^{5}$ITA, Albert-$\ddot{\rm U}$berle-Strasse 2, 69120 Heidelberg, Germany}

\maketitle
\begin{abstract}
We present the first tests of a new method, the Correlated 
Component Analysis (CCA) based on second-order statistics, to 
estimate the mixing matrix, a key ingredient to separate astrophysical 
foregrounds superimposed to the Cosmic Microwave Background (CMB). 
In the present application, the mixing matrix is 
parameterized in terms of the spectral indices of
Galactic synchrotron and thermal dust emissions, while the
free-free spectral index is prescribed by basic physics, and is
thus assumed to be known. We consider simulated observations of
the microwave sky with angular resolution and white stationary
noise at the nominal levels for the {\sc Planck} satellite, and
realistic foreground emissions, with a position dependent
synchrotron spectral index. We work with two sets of {\sc Planck}
frequency channels: the low frequency set, from 30 to 143 GHz,
complemented with the Haslam 408 MHz map, and the high frequency
set, from 217 to 545 GHz. The concentration of intense free-free
emission on the Galactic plane introduces a steep dependence of
 the spectral index of the global Galactic emission with Galactic
latitude, close to the Galactic equator. This feature makes
difficult for the CCA to recover the synchrotron spectral index in
this region, given the limited angular resolution of {\sc Planck},
especially at low frequencies. A cut of a narrow strip around the
Galactic equator ($|b|<3^\circ$), however, allows us to overcome
this problem. We show that, once this strip is removed, the 
CCA allows an effective foreground subtraction, with residual 
uncertainties inducing a minor contribution to errors on the recovered 
CMB power spectrum. 

\end{abstract}
\section{Introduction}
The Cosmic Microwave Background (CMB) is by far the most powerful
cosmological probe. The power spectra of its temperature and
polarization anisotropies encode detailed information on the key
cosmological parameters. Tremendous experimental efforts, and
especially the currently flying \emph{Wilkinson Microwave
Anisotropy Probe} (WMAP, Bennett et al. 2003a, Spergel et al. 2006) 
and the forthcoming 
{\sc Planck} satellite (Tauber 2004; Lamarre et al. 2003;
Mandolesi, Morgante, \& Villa 2003) will substantially advance the
sensitivity and resolution of maps of the microwave sky.
Correspondingly, efficient methods to extract all the available
information from the measured signal need to be implemented,
exploiting the advances of the data analysis science. One of the
most challenging tasks is the separation of the astrophysical
components superimposed on the CMB, usually denominated
``foregrounds''.

On angular scales larger than about $30'$ the dominant foregrounds
in the relevant spectral region are diffuse emissions from our own
Galaxy (De Zotti et al. 1999). Synchrotron (Haslam et al. 1982)
and free-free (Haffner, Reynolds, \& Tufte 1999, Finkbeiner 2003)
emissions dominate below $\simeq 60-80$ GHz (Bennett et al.
2003b), while at higher frequencies thermal dust (Schlegel,
Finkbeiner, Davies 1998; Finkbeiner, Schlegel, Davies 1999) takes
over. On smaller angular scales foreground fluctuations are
dominated by several populations of extra-galactic sources, with
different spectral behaviour: radio sources, dusty galaxies and
the Sunyaev-Zel'dovich effect from galaxy clusters.

A big deal of work has been recently dedicated to develop
algorithms performing component separation, based on different
ideas and techniques from signal processing science. Several
algorithms, referred to as ``non-blind'', assume a perfect
knowledge of the frequency dependence of sources. The most widely
used techniques exploiting this approach are Wiener Filtering (WF;
Tegmark \& Efstathiou 1996; Bouchet, Prunet, \& Sethi 1999) and
the Maximum Entropy Method (MEM; Hobson et al. 1998; Barreiro et
al. 2004; Maisinger et al. 2004; Stolyarov et al. 2005). On the
other hand, the emission spectra are generally poorly known. This
fact has motivated ``blind'' approaches, which do not make any
assumption on the spectral shape. Particularly promising are
methods exploiting Independent Component Analysis (ICA) techniques
(Amari \& Chichocki 1998; Hyv\"arinen 1999), relying on the
statistical independence of the different components. These
methods proved to be very effective in extracting the CMB, which
is independent of all other sources (Baccigalupi et
al. 2000, 2004; Maino et al. 2002, 2003; Stivoli et al. 2006; 
Patanchon et al. 2005).
However, Galactic
emissions are tightly correlated to each other, so that the assumption of
mutual independence breaks down.
Still following the ICA approach, Belouchrani 
et al. (1997) dropped statistical independence for mere uncorrelation,
and recovered the missing information by exploiting the physically
plausible assumption of a significant spatial autocorrelation of the
individual sources. 
Bedini et al. (2005) also exploited second-order statistics and spatial
autocorrelation but, at variance with Belouchrani et al. (1997), they
gave up the assumption of mutually uncorrelated sources. By exploiting a
parametric knowledge of the mixing matrix, they succeeded in estimating both the
mixing matrix and the relevant cross-covariances.

Here we present the first tests on this method, referred to as
\emph{correlated component analysis} (CCA), on a data set as realistic as
possible. Also, we work either on the whole sky or on spherical sky
patches, rather than on small plane patches as in Bedini et al. (2005).

The paper is organized as follows. In Section 2 and 3 
we describe the basic aspects of CCA, and its implementation. 
In Section 4 we illustate the tests on simulated skies. 
In Section 5 and 6 we estimate the precision of the mixing matrix 
and CMB power spectrum recovery. In Section 7 we draw our conclusions. 

\section{The Correlated Component Analysis (CCA)}\label{sec:cca}
We assume that the observed sky radiation is the superposition of
$N$ different physical processes whose spatial pattern is
independent of frequency spectrum:
\begin{equation}
x(\mathbf{r},\nu)=\sum_{j=1}^{N}s_j(\mathbf{r})f_j(\nu).
\end{equation}
If we have a set of equal resolution observations at $M$ different
frequencies, the observed signal can be modelled as:
\begin{equation}
\mathbf{x}=\mathbf{H}\mathbf{s}+\mathbf{n} \label{model},
\end{equation}
where $\mathbf{x}$ is the $M$-vector of observations, $\mathbf{H}$
is a $M \times N$ mixing matrix, $\mathbf{s}$ is the $N$-vector of
sources and  $\mathbf{n}$ the $M$-vector of instrumental noise.
The generic element of the mixing matrix is related to the source
spectra, $f_c(\nu)$, and to the instrumental frequency response
function, $b_d(\nu)$:
\begin{equation}
h_{dc}=\int f_c(\nu) b_d(\nu)d\nu \ . \label{hdc}
\end{equation}
Assuming that the source spectra are constant within the
passbands, eq.~(\ref{hdc}) becomes:
\begin{equation}
h_{dc}=f_c(\nu_d)\int b_d(\nu)d\nu ,\label{hdc2}
\end{equation}
meaning that the element $h_{dc}$ is proportional to the spectrum
of the c-th source at the  central frequency $\nu_d$ of the d-th
channel.
In any case, $h_{dc}$ will be  
proportional
to $f_c$ at an effective frequency $\nu_{\rm eff}$ within the $d$-th  
sensor passband,
depending on both the spectrum and the frequency response (see Eriksen
et al. 2006).

If both the mixing matrix $\mathbf{H}$ and the source vector
$\mathbf{s}$ are unknown, the problem is unsolvable without
additional hypotheses. The CCA method exploits information on the
second order statistics of the data.

The covariance matrix of a generic signal $\mathbf{X}$, defined in
a two dimensional space with coordinates $(\xi,\eta)$, is:
\begin{eqnarray}
\mathbf{C_x}(\tau,\psi)=\langle [\mathbf{X(\xi,\eta)}-\mu]
[\mathbf{X}(\xi+\tau,\eta+\psi)-\mu]^T \rangle \label{covmat},
\end{eqnarray}
where $\langle ...
\rangle$ denotes expectation under the appropriate joint
probability distribution, $\mu$ is the mean vector and the
superscript $T$ means transposition. Every covariance matrix is
characterized by the shift pair $(\tau,\psi)$, where $\tau$ and
$\psi$ are increments in the $\xi$ and $\eta$ coordinates.

From eq.~(\ref{model}) we can easily derive a relation between the
data covariance matrix $\mathbf{C_x}$ at a certain lag, the source
covariance matrix $\mathbf{C_s}$ at the same lag, the mixing
matrix $\mathbf{H}$, and the noise covariance matrix
$\mathbf{C_n}$:
\begin{equation}
\mathbf{C_x}(\tau,\psi)=\mathbf{H}\mathbf{C_s}(\tau,\psi)
\mathbf{H}^T+\mathbf{C_n}(\tau,\psi).\label{problem}
\end{equation}
This relation allows us to estimate the mixing operator
$\mathbf{H}$ from the covariance matrix $\mathbf{C_x}$ of the
data. Note that it implicitly assumes that the source and the
noise processes are stationary. This is true all across the sky
for the CMB, but only within small sky patches for the foregrounds.
For this reason, it is convenient to section the sky into patches
within which foregrounds have approximately uniform properties,
and apply the method to individual patches. 
In this way the only remaining issue is the non-stationarity of noise, 
depending on the particular scanning strategy adopted, which we do not 
consider in this work. 

If the noise process can be assumed to be signal-independent,
white and zero-mean, for $(\tau,\psi)=(0,0)$ $\mathbf{C_n}$ is a
diagonal matrix whose elements are the noise variances in the
frequency channels of the instrument, while for  $(\tau,\psi)\neq
(0,0)$ $\mathbf{C_n}$ is the null $M \times M$ matrix. Anyway, if
$\mathbf{C_n}$ deviates significantly from this ideal model,
various methods are available to estimate the noise covariance
function: for example it can be empirically determined using noise 
maps from Monte Carlo simulations. 

Once we have a model for $\mathbf{C_n}$, we only have to calculate
$\mathbf{C_x}$ for a large enough number of nonzero shift pairs
$(\tau,\psi)$ to estimate both $\mathbf{H}$ and $\mathbf{C_s}$. In
practice, however, we need to parameterize the mixing matrix to
reduce the number of unknowns. We note that for the scaling
ambiguity, we can normalize our mixing matrix to have all elements
of a reference row equal to 1.

The main product of CCA is then an estimate of the mixing matrix.
Hence this can be considered a ``model learning'' algorithm. Once
the mixing matrix have been recovered, source separation can be
performed with traditional non-blind methods, such as WF or MEM,
or other Bayesian inversion techniques.

\subsection{Parameterization of the mixing matrix}

To choose a suitable parameterization for $\mathbf{H}$ we use the
fact that its elements are proportional to the spectra of
astrophysical sources, of which we have some knowledge coming from
the theory or from complementary observations. The main diffuse
components present in the {\sc Planck} channels are, in addition
to the CMB, the Galactic dust, synchrotron and free-free
emissions. The frequency dependencies of the CMB and of the
free-free are known; in terms of the antenna temperature we have:
\begin{equation}
T_{\rm A,CMB}(\nu)\propto\frac{(h\nu/kT_{\rm CMB})^2\exp (h\nu/kT_{\rm CMB})}
{(\exp (h\nu/kT_{\rm CMB})-1)^2} \label{cmb}\ ,
\end{equation}
\begin{equation}
T_{\rm A,ff}(\nu)\propto (\nu)^{-2.14} \label{ff},
\end{equation}
where $h$ is the Planck constant, $k$ the Boltzmann constant,
$T_{\rm CMB}=2.726$ K. Conversely, the frequency scalings of
synchrotron and dust are not known a priori. The simplest model
compatible with observations involves only one parameter for each
source, the spectral index $\beta_s$ or the emissivity index
$\beta_d$, respectively:
\begin{equation}
T_{\rm A,synch}(\nu)\propto \nu^{-\beta_s} \label{syn}\ ,
\end{equation}
\begin{equation}
T_{\rm A,dust} (\nu)\propto \frac{\nu^{\beta_d+1}}{\exp
(h\nu/kT_{\rm dust})-1} \label{dust}.
\end{equation}
The mixing matrix accounting for all these sources is then of four
columns and of as many rows as the number of channels. 
We are able to parameterize the matrix $\mathbf{H}$ by exploiting
eqs.~(\ref{hdc}) or (\ref{hdc2}). The integrals in eq.~(\ref{hdc}) can be evaluated
with $f_c$ replaced by one of the emission spectra in eqs.~(\ref{cmb})-(\ref{dust}).

Under the
above assumptions, we only have two free parameters, $\beta_s$ and
$\beta_d$. When we work with the high frequency {\sc Planck}
channels, synchrotron and free-free can be neglected, so the
mixing matrix has two columns, one for the CMB and one for the
dust, and only one parameter, $\beta_d$.

Since $\beta_s$ and $\beta_d$ vary across the sky, we will apply
the CCA to sky patches small enough for these indices can be
assumed to be approximately constant, as we specify in Section 5. 
\section{Implementation}
\subsection{Computation of the data covariance matrix}
The data covariance matrix defined by eq.~(\ref{covmat}) is the
fundamental tool used to identify the mixing operator. If our data
are sampled in $P$ pixels, labelled with coordinates $(\xi,\eta)$,
we can compute an estimate of $\mathbf{C_x}$ as:
\begin{eqnarray}
\hat{\mathbf{C_x}} (\tau,\psi) =&{1\over P} \sum_{\xi, \eta}
[\mathbf{x}(\xi,\eta)-\mu _x ]\cdot \nonumber \\
 & \cdot [\mathbf{x}(\xi+\tau,\eta+\psi)-\mu _x ]^T \label{covariance}.
\end{eqnarray}
The shift pair $(\tau,\psi)$ defines a vector that links each
pixel to a shifted one. In the case of a uniformly sampled
rectangular sky patch we have a very easy way to define the shift
pairs $(\tau,\psi)$. We choose $(\xi,\eta)$ as cartesian
coordinates whose axes are parallel to the sides of the rectangle,
and define a collection of $N_p$ shifts, each of $p$ pixels,
labelled by an index $n$ ($n=0,1,..N_p-1$):  $\{\tau_n =n p
\mathbf{u_\xi} \}$ and $\{\psi_n =n p \mathbf{u_\eta} \}$. From
all the combinations of $\tau_n $ and $\psi_n $ we get  $N_p
\times N_p$ shift pairs.

The data we are working with are very different from this simple
case: they are patches extracted from Healpix (G\'orski et al.
2005) 
all sky maps, so they are sampled in a sphere rather than in
a plane and also the grid is not regular. We note that even in the
simplest case the selected region is not exactly rectangular
because of the diamond shape of pixels and of the surface
curvature.

To apply the method we need to map the selected patch into a
geometrically identical one, shifted by $(\tau,\psi)$. This can be
done only in an approximate way with the Healpix pixelization, and
is easier in the equatorial region. In the present application we
study Galactic foregrounds where they are more intense, i.e. at
not too high Galactic latitudes. We therefore use Galactic
coordinates and refer to the grid defined by Galactic parallels
and meridians to calculate shifts.

With the Healpix ring ordering scheme, pixels with subsequent
indices are subsequent in longitude, so that, for any integer $p$,
a set of pixels $i_{\rm min} \le i \le i_{\rm max}$ simply map
into the set $i_{\rm min}+p \le i \le i_{\rm max}+p$,  shifted by
$p$ pixels in the longitudinal direction. The shift in latitude is
trickier. We proceeded associating to each pixel the shifted one
closest to having the same longitude and the latitude increased by
$\Delta b= p\cdot ds$, where $ds$ is the mean pixel size. Clearly,
in this case $\Delta b$ is not exactly equal for all shifted
pixels. However, our shifts are rather small (see below), so that,
if we are not too close to the Galactic poles, the approximation
is sufficiently good.

To choose convenient values for the number of shifts $N_{p}$ and the
step $p$, we should care both about efficiency and conditioning. It is
obvious that, if $N_{p}$ is large, the method becomes computationally 
demanding. On the other hand, a too small $N_{p}$ can make
the problem ill conditioned, thus leading to a lack of convergence. The
step $p$ should be chosen in order to avoid both redundancy (if $p$ is
small, some of the covariance matrices are nearly equal) and degeneracy
(if $p$ is too large, some covariance matrices vanish). In practice,
the choice can be made empirically, for any $p$, by increasing $N_{p}$
progressively until convergence is reached.

\subsection{The minimization procedure}
To solve eq.~(\ref{problem}) and estimate the parameters that identify
the mixing matrix and the source covariance matrices, we minimize the 
residual between the theoretical quantities based on the proposed
solution and the corresponding quantities evaluated empirically from
the available data. Our solution is given by:
\begin{eqnarray}
(\Gamma,\Sigma(:,:))\!\!\!\!\!\!&=&\!\!\!\!\!\! {\rm argmin} \sum
_{\tau,\psi}\parallel \mathbf{H}(\Gamma)\mathbf{C_{s}}[
\Sigma(\tau,\psi)]
\mathbf{H}^T(\Gamma)- \nonumber \\
& &- \mathbf{\hat{C}_{x}}(\tau,\psi)+ \mathbf{C_{n}} (\tau,\psi)
\parallel \label{objective},
\end{eqnarray}
where $\Gamma$ is the vector of all parameters defining
$\mathbf{H}$, and $\Sigma(:,:)$ is the vector containing all the
unknown elements of the matrices $\mathbf{C_{s}}$ 
for every shift pair.

To perform the minimization, we used simulated annealing 
(SA, Aarts \& Korst, 1989), which exploits an analogy between the way in which a
metal cools and freezes into a minimum energy crystalline
structure and the search for a minimum in a more general system.

The major advantage of SA over other methods is its ability to
avoid becoming trapped at local minima, which can be very nasty in
our case. The algorithm employs a random search which not only
accepts changes that decrease the objective function $f$, but also
some changes that increase it. The latter are accepted with a
probability:
\begin{equation}
p=\exp(-\delta f/T),
\end{equation}
where $\delta f$ is an increment in $f$ and $T$ is a control
parameter, known as the system ``temperature''.

Avoidance of local minima is anyway dependent on the ``annealing
schedule'': the choice of the initial temperature, how many
iterations are performed at each temperature, and how much the
temperature is decreased at each step as cooling proceeds.

\section{Tests on simulated skies}\label{sec:simulations}

\begin{table*}
\caption{{\sc Planck} specifications. We assume spatially uniform
Gaussian noise at the mean level expected for the nominal mission
(14 months)} \label{tab:planck}

\begin{tabular}{|l|l|l|l|l|l|l|l|l|l|}
\hline
 Center frequency (GHz)&30&44&70&100&143&217&353&545&857\\
\hline
 Angular resolution (arcmin)&33&24&14&9.5&7.1&5.0&5.0&5.0&5.0\\
\hline
 Rms pixel noise $\Delta T$ ($\mu$K thermodynamic)&5.5&7.4&12.8&6.8&6.0&13.1&40.1&401&18291\\
\hline
\end{tabular}
\end{table*}

For this set of tests we used the specifications of the {\sc Planck}
mission (see Table \ref{tab:planck}).  The simulated sky contains: a)
synchrotron emission as modelled by Giardino et al.  (2002), allowing
for a spatially varying spectral index; b) thermal dust emission
(Finkbeiner et al.  1999), with dust at two temperatures and two
emissivity indices; c) free-free traced by the $H_\alpha$ emission
(Dickinson et al.  2003) and corrected for dust absorption with the
100 $\mu$m maps from Schlegel et al. (1998); d) a CMB Gaussian
realization corresponding to the best fit WMAP
theoretical power spectrum from first year data. 
We produced 100 sets of Monte Carlo simulated maps in the {\sc
Planck} channels, with different realizations of the CMB and of
Gaussian noise.

To test the CCA ability to recover the spectral parameters of
foregrounds, we used the {\sc Planck} channels from 30 to 143 GHz.
The resolution of all the maps had to be degraded to that of the
30 GHz channel ($33'$). To test the performances achievable with
the full {\sc Planck} resolution of $5'$, we repeated the analysis
with the high frequency channels from 217 to 545 GHz, where the
sources to account for are only CMB and dust.

The noise maps we initially added to each channel were Gaussian (with
the rms levels reported in Table\ref{tab:planck}) and uncorrelated: in
this case, the only noise term in eq.~(\ref{problem}) is
$\mathbf{C_n}(0,0)$, which is a diagonal matrix whose elements are the
noise variances for each channel.  The smoothing process applied to
degrade the channels to the 30 GHz resolution not only changes the
noise variances, but also introduces noise correlation, so in
principle the terms $\mathbf{C_n}(\tau,\psi)$ do not vanish for
$(\tau,\psi)\neq(0,0)$.  Nevertheless, we carried out our tests
assuming uncorrelated Gaussian noise: we then consider only the
diagonal matrix $\mathbf{C_n}(0,0)$, whose elements are the variances
measured after smoothing each noise map to the $33'$ resolution.  This
approximation is not the best we could do in dealing with noise, but
it was purposely adopted to investigate the effect of errors on noise
modelling.

\subsection{The low frequency channels (LF) set }

The LF set includes 5 {\sc Planck} channels, centered at 30, 44,
70, 100, and 143 GHz. Since we have 4 sources (CMB, synchrotron,
free-free, and thermal dust) the mixing matrix has 4 columns and
we want to recover the synchrotron and the dust spectral indices.
Since all the maps have been degraded to a resolution of $33'$,
we operate with the Healpix parameter NSIDE=512, corresponding 
to a pixel size of about 7'. 

The number of shifts allowing a good conditioning of the problem
is found to be $N_p=5$. The value of $N_p$ and the pixel size
constrain the amplitude of the shifts that can be used to
calculate the covariance matrices: the minimum shift must
correspond to one pixel ($\sim 0.1 ^\circ$); the maximum one
cannot generally exceed $1^\circ$ to ensure that all covariance
matrices are non-null. We chose $p=4$, which corresponds roughly
to our beam size.

There is a second effect of the resolution: to have sufficient
statistics, the number of pixels per patch has to be at least
$\simeq 10^5$. With the adopted pixel size, this corresponds to a
patch area of $1500\,\hbox{deg}^2$, which is not optimal for the
reconstruction of the spectral indices that may vary widely with
the position. In the synchrotron template we use here (Giardino 
et al. 2002), $10\%$ variations occur on scales of about 
10 degrees. 

We used patches of $(\Delta l,\Delta b)=(50^\circ, 30^\circ)$ for
the lowest latitudes and increased the longitudinal dimension for
higher latitudes to roughly preserve the patch area. We
sectioned the sky with patches centered at longitudes $l_c
=\{0^\circ,40^\circ, 80^\circ,... 320^\circ \}$ and increasing
latitudes.  In total we analyzed the $\sim 80\%$ of the sky with
$|b|<55^\circ$.

From the first tests we learned that to get a good spectral index
reconstruction we need a broad frequency range. This can be
achieved by taking into account additional information from other
surveys. To this end, we included in our analysis the 857 GHz {\sc
Planck} map and the Haslam et al. (1982) 408 MHz map, taken as
dust and synchrotron templates, setting to 0 all the elements of
the mixing matrix except the synchrotron one at 408 MHz, and the
dust one at 857 GHz.

\subsection{The high frequency channels (HF) set}
In the {\sc Planck} 217, 353 and 545 GHz channels we can neglect the
synchrotron and free-free emissions, so that we are left with CMB
and dust, and the only parameter to recover is the dust emissivity
index $\beta_d$. These channels allow us to work at the best {\sc
Planck} resolution: the beam is of $5'$, so we can use NSIDE=2048, 
corresponding to a pixel size of about 1'.7. 
We dissected the sky into patches of size $(\Delta l,\Delta b)=(20^\circ, 20^\circ)$, 
centered at longitudes $l_c =\{0^\circ,20^\circ, 40^\circ,...
340^\circ \}$ and latitudes $b_c=\{0^\circ,\pm10^\circ,
\pm20^\circ, \pm40^\circ ,\pm60^\circ\}$. In total, we then
analyzed $\sim 85\%$ of the sky.

In this case, we did not need to include other channels to help
reconstructing the mixing matrix.  We used a shift step $p=5$,
corresponding to an angular size of $0.14^\circ$, and $N_p=3$.

\section{Estimation of the mixing matrix}

\subsection{Error estimates}\label{sec:overline}

Since, on purpose, the parameters to be determined do not directly
reflect those defining the sky model, the errors on our estimates
cannot be simply derived comparing our results with the ``true''
values, simply because the latter in general do not exist. In particular, 
the synchrotron spectral index is varying in the sky,
and the dust emission is modeled by a two-component spectrum, so they need to be treated 
as we describe below. 

The output of our method is the estimated mixing matrix
normalized at a reference frequency $\nu _0$. The elements of the
mixing matrix corresponding to synchrotron and dust at
frequency $\nu_d$ are:
\begin{equation}
h_{\rm out}({\rm syn}, \nu_d)=\Big(\frac{\nu_d}{\nu_0}\Big)^{-\beta_s},
\label{houtsyn}
\end{equation}
\begin{equation}
h_{\rm out}({\rm dust},
\nu_d)=\Big(\frac{\nu_d}{\nu_0}\Big)^{\beta_d+1}
\frac{\exp(h\nu_0/kT_{\rm dust})-1}{\exp(h\nu_d/kT_{\rm dust})-1}
\label{houtdust},
\end{equation}
which were derived from eqs.~(\ref{hdc}), (\ref{syn})
and (\ref{dust}) by assuming $b_d(\nu)=\delta(\nu - \nu_d)$.
These quantities correspond to the mean ratios of synchrotron and
dust intensities at frequencies $\nu_0$ and $\nu_d$, within the
considered patch in the simulated sky:
\begin{equation}
h_{\rm in}({\rm syn},
\nu_d)=\Big(\frac{\nu_d}{\nu_0}\Big)^{-\overline{\beta}_s},
\label{houtsyn2}
\end{equation}
\begin{equation}
h_{\rm in}({\rm dust},
\nu_d)=\Big(\frac{\nu_d}{\nu_0}\Big)^{\overline{\beta}_d+1}
\frac{\exp(h\nu_0/kT_{\rm dust})-1}{\exp(h\nu_d/kT_{\rm dust})-1}
\label{houtdust2}.
\end{equation}
The ``observed'' indices  $\overline{\beta}_s$ and 
$\overline{\beta}_d$ are now directly comparable with the derived 
indices $\beta_s$ and $\beta_d$. We then define our errors as: 
\begin{equation}
\Delta \beta_s=\Big|\frac{\log (h_{\rm out}({\rm syn}, \nu_d)/h_{\rm in}({\rm syn}, \nu_d))}
{\log(\nu_d /\nu _0)}\Big|\label{errsyn1}\ ,
\end{equation}
\begin{equation}
\Delta \beta_d=\Big|\frac{\log (h_{\rm out}({\rm dust},
\nu_d)/h_{\rm in}({\rm dust}, \nu_d)}{\log(\nu_d /\nu _0)}\Big|
\label{errdust1}.
\end{equation}

\subsection{Results for the mixing matrix identification}

\begin{figure*}
\begin{center}
\includegraphics[width=11cm, angle=90.]{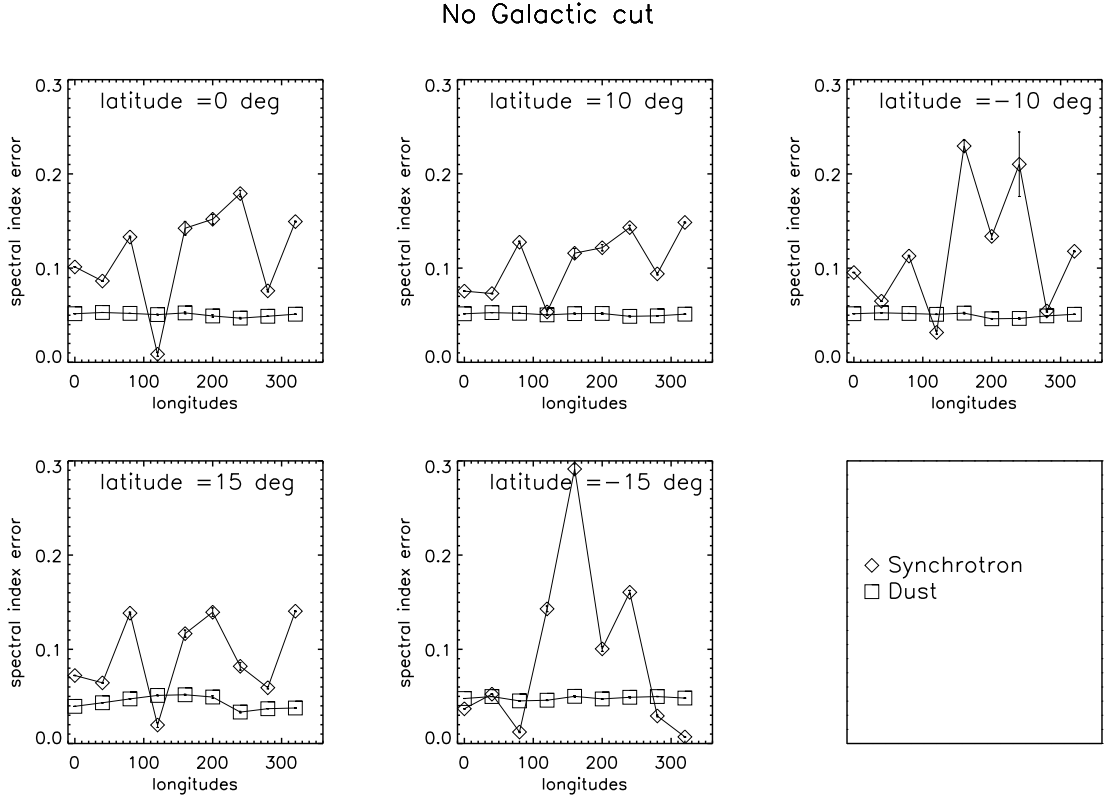}
\caption{Absolute errors in the spectral index reconstruction for the LF
channels without cuts} \label{fig:uno}
\end{center}
\end{figure*}

\begin{figure*}
\begin{center}
\includegraphics[width=11cm, angle=90.]{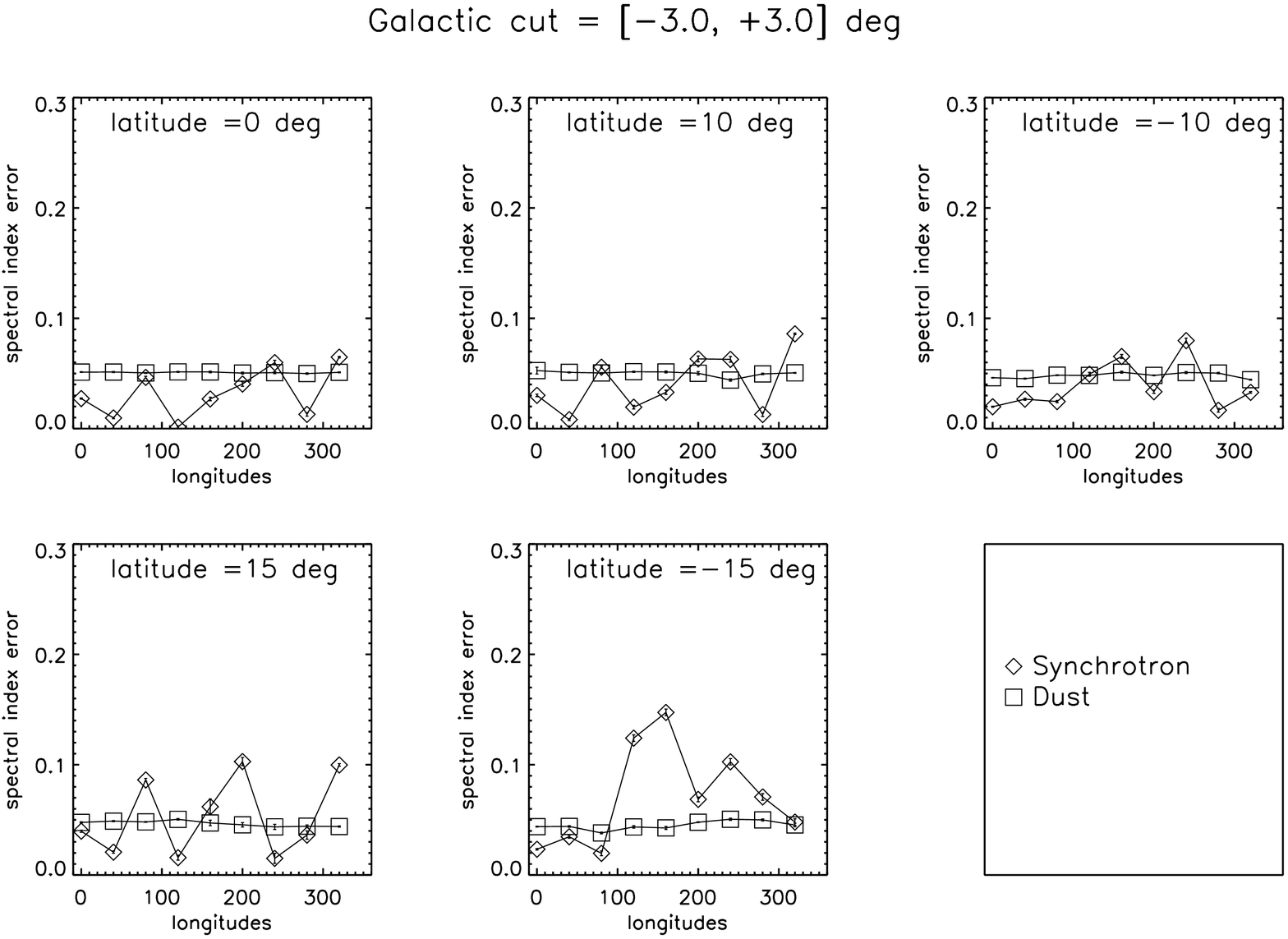}
\caption{Absolute errors in the spectral index reconstruction for  LF
channels cutting out the Galactic plane region $|b|\le 3^\circ$}
\label{fig:due}
\end{center}
\end{figure*}

\begin{figure*}
\begin{center}
\includegraphics[width=6cm, angle=90.]{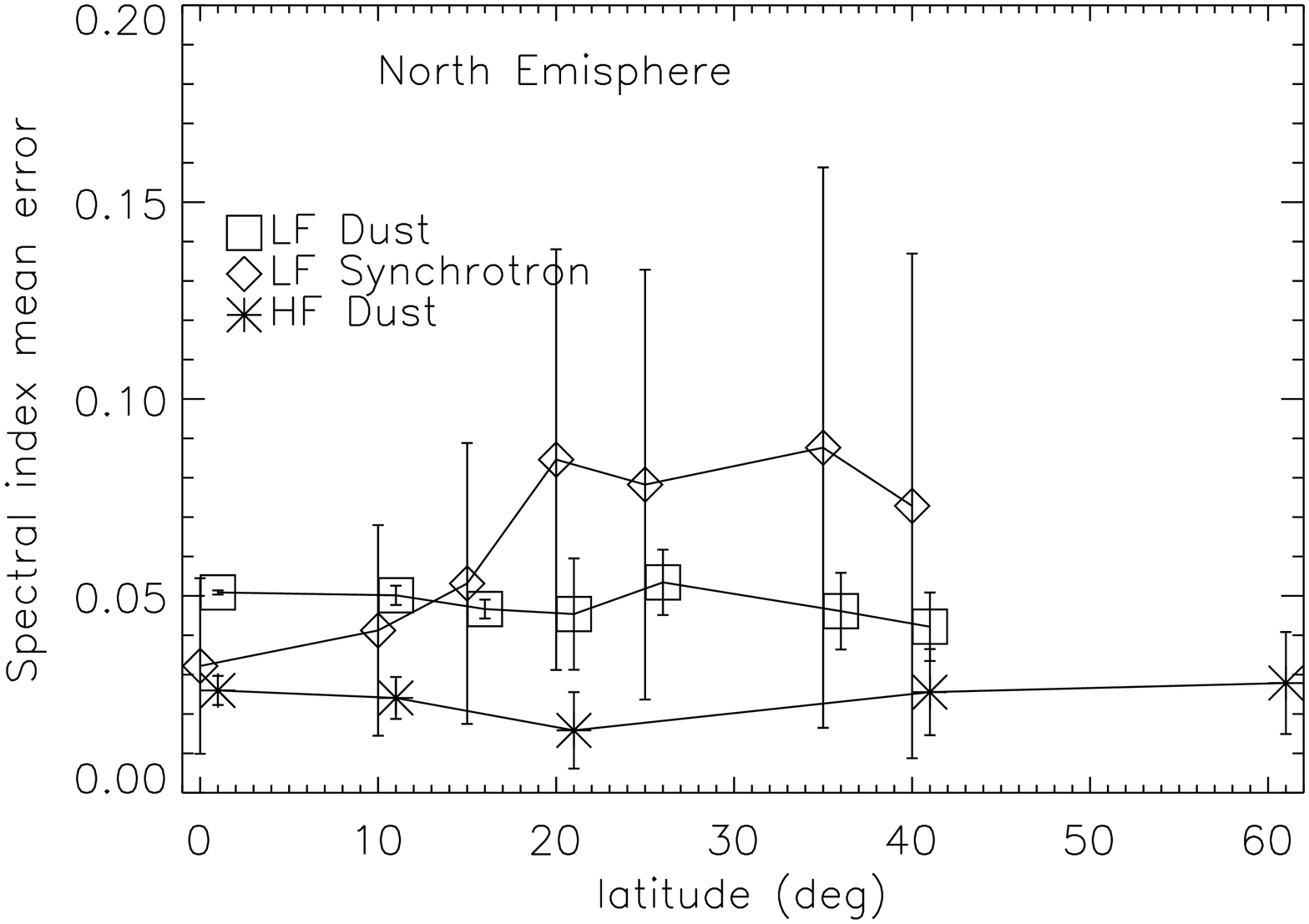}
\includegraphics[width=6cm, angle=90.]{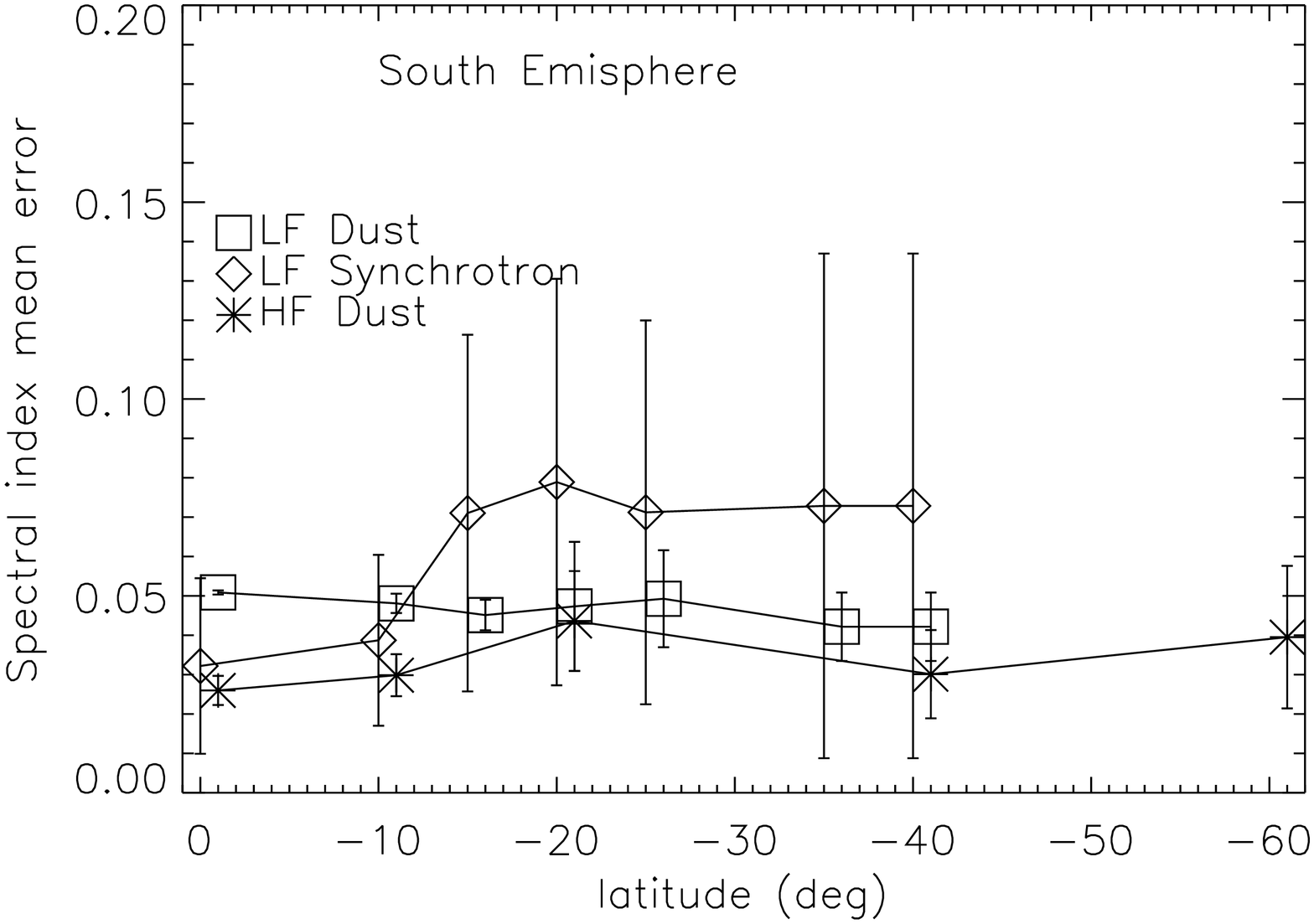}
\caption{Mean absolute spectral index errors versus Galactic latitude 
for the analysis of LF and HF channels with the Galactic cut of $\pm
3^\circ$} \label{fig:errvslat}
\end{center}
\end{figure*}

Figure~\ref{fig:uno} refers to the analysis of the LF set and 
shows the mean values of $\Delta \beta_s$ and $\Delta \beta_d$,
over 100 simulations of the sky, as a function of longitude for
latitudes $b=\{0^\circ, \pm 10^\circ, \pm 15^\circ\}$. 
The error bars, generally very small and barely visible in the figure, 
are the standard deviations of $\Delta \beta_s$ and $\Delta
\beta_d$ from the mean. Their small values imply that the errors 
in the spectral index recovery are mainly systematic; for both spectral 
indices, part of the systematic error comes from the 
fact that the models we assume for the analysis differ from those defining the sky model.
We clearly have problems with the 
estimation of the synchrotron spectral index, while the errors on 
$\beta_d$ are generally small.

We have checked that this effect is due to confusion between
synchrotron and free-free. The free-free emission is highly
concentrated on the Galactic plane but depresses the mean spectral
index of the combined emission over entire patches, thus biasing the
estimate of the synchrotron spectral index. A better recovery of
the latter would be possible with high spatial resolution maps, so
that we have a sufficient number of pixels within a much smaller
range of Galactic latitudes.

With the angular resolution adopted here, this effect can be
minimized by cutting out a strip of a few degrees around the Galactic
plane. The situation is substantially better even with a cut of
$\pm 1^\circ$ around the Galactic equator, and improves further if
we enlarge the cut to $\pm 3^\circ$ (Fig.~\ref{fig:due}). All the
tests described below are performed by applying this
cut, even though its effects are modest when working with
the HF channels.

Figure {\ref{fig:errvslat}} summarizes the results obtained over
all the analyzed region of the sky with both the LF and HF sets. 
We computed the mean errors on synchrotron and dust indices over 100
Monte Carlo iterations for each patch, and the mean and standard
deviation of errors obtained for patches at the same latitude. The
bars are the standard deviations around the mean over longitudes.

For the LF set, while the dust index is always reconstructed with
an error of $\simeq 0.05$, the error on the synchrotron spectral
index increases with latitude. This is due to the fact that the
substantial spatial variability of the synchrotron spectral index,
implied by the Giardino et al. (2002) model, is increasingly
difficult to recover as the synchrotron signal weakens with
increasing Galactic latitude.

Thanks to the higher angular resolution of the HF set and to the
lack of other relevant diffuse components besides CMB and dust in
this frequency range, the CCA allows us to reconstruct the dust
spectral index with an error $\sim 0.03$ over the full latitude
range analyzed.

\section{Estimation of errors in the CMB power spectrum}

The CCA procedure can be viewed as the first step (learning) in
component separation.  In a further step, the mixing matrix estimated
by this approach can be input to non-blind separation tecniques.  The
performances of non-blind techniques are normally evaluated as
functions of the system noise, assuming perfectly known mixing
matrices.  In this section, we estimate the uncertainties on the CMB
power spectrum induced by the errors in the mixing matrix resulting
from our simulations.  Since at the moment the optimal component
separation method exploiting the information recovered by CCA has still to
be developed, we try to estimate the errors in the CMB spectrum under conservative hypotheses.  Clearly, the errors will
depend on the approach adopted to separate the individual components.
We investigated the general case where component separation is
performed by a generic linear filter and, in particular, by a Wiener
filter, and a pseudo-inverse reconstruction.  The data model we assume
provides a space-invariant mixing matrix $\mathbf{H}$, obtained with spectral indices $\overline{\beta}_s$ and $\overline{\beta}_d$ as defined in Section \ref{sec:overline}. To evaluate the errors caused by an
approximated estimate of the mixing matrix, we assumed the estimated
matrix, $\mathbf{\hat{H}}$, to be generated by a distribution of the
spectral indices such as the one described in the above sections. Following a theoretical derivation, the errors are evaluated in terms of
quantities that are known for our simulation, thus allowing the
estimation to be made without actually separating the components.

Given the mixing matrix, one way to solve eq.~(\ref{model}) is to
work in a conjugate space.  If we work with sky patches that
are small enough for curvature effects can be neglected, the correct basis
functions are the Fourier complex exponentials; on the whole celestial
sphere, the good basis functions are the spherical harmonics; finally,
in the presence of an incomplete sky coverage, pixelization, and
position-dependent noise, the basis functions can be calculated
according to Tegmark (1996).  For simplicity, we are going to derive
the equations in spherical harmonics.

In the harmonic space, the problem stated in eq.~(\ref{model}) simply becomes:
\begin{equation}
\mathbf{x}_{lm}=\mathbf{H}\mathbf{s}_{lm}+\mathbf{n}_{lm} \label{model2},
\end{equation}
where the vectors $\mathbf{x}_{lm}$, $\mathbf{s}_{lm}$, and
$\mathbf{n}_{lm}$ contain the harmonic coefficients of channels,
sources and instrumental noise, respectively. 
Using a linear approach to component
separation, an estimate of the vector $\mathbf{s}_{lm}$, say $\mathbf{\hat{s}%
}_{lm}$, can be obtained as: 
\begin{equation}
\mathbf{\hat s}_{lm}=\mathbf{W}_{\hat{H}}^{(l)}\mathbf{x}_{lm}\label{recon},
\end{equation}
where $\mathbf{W}_{\hat{H}}^{(l)}$, referred to as the
\emph{reconstruction matrix}, is some matrix-valued function depending
on the estimate $\mathbf{\hat{H}}$ of the mixing matrix.  The random
matrix $\mathbf{\hat{H}}$ is estimated with the method proposed in this
paper, and is a function of the estimated spectral indices.  
Following Tegmark \& Efstathiou (1996) and Bouchet \& Gispert (1999),
the estimation error on the source spectra, in matrix notation, can be
expressed as:
\begin{equation}
\Delta \mathbf{C}_{\hat{H}}^{(l)}=\frac{1}{2l+1}\sum_{m=-l}^{l}\langle
\left( \mathbf{\hat{s}}_{lm}-\mathbf{s}_{lm}\right) \left( \mathbf{\hat{s}}%
_{lm}-\mathbf{s}_{lm}\right) ^{\dagger }\rangle  \label{22},
\end{equation}%
where $\langle...\rangle $ indicates expectation and 
$\dagger $ indicates conjugate transposition. 
By using eqs.~(\ref{model2}) and
(\ref{recon}),
after a straightforward calculation, we have:
\begin{eqnarray}
\Delta \mathbf{C}_{\hat{H}}^{(l)} &=&  ( \mathbf{W}_{\hat{H}}^{(l)}\mathbf{%
H-I}) \mathbf{C_{s}}^{(l)}( \mathbf{W}_{\hat{H}}^{(l)}\mathbf{H-I}%
) ^{\dagger } + \nonumber \\ &+& \mathbf{W}_{\hat{H}}^{(l)}\mathbf{C_{n}}^{(l)} ( 
\mathbf{W}_{\hat{H}}^{(l)}) ^{\dagger },  \label{23}
\end{eqnarray}%
where $\mathbf{I}$ is the identity matrix, and $\mathbf{C_{s}}$ and
$\mathbf{C_{n}}$ are, respectively, the source and the noise power
spectra.

Eq.~(\ref{23}) can be simplified by assuming that the source and the noise processes are mutually uncorrelated: in this case  $\mathbf{C_{s}}$,
$\mathbf{C_{n}}$ and $\Delta \mathbf{C}_{\hat{H}}^{(l)}$ are diagonal matrices. If we define:
\begin{eqnarray}
C_{s}^{(l)}(i) &\equiv &\mathbf{C_{s}}^{(l)}(i,i)\ ,\\
C_{n}^{(l)}(i) &\equiv &\mathbf{C_{n}}^{(l)}(i,i),
\end{eqnarray}
the estimation error on the $i$-th source is:
\begin{eqnarray}
\Delta C_{\hat{H}}^{(l)}(i) &\equiv& \Delta \mathbf{C}_{\hat{H}}^{(l)}(i,i)= \sum_{j=1}^{N}|(\mathbf{W}_{\hat{H}}^{(l)}%
\mathbf{H-I})_{ij}|^{2}C_{s}^{(l)}(j)+ \nonumber \\ &+&\sum_{j=1}^{M}|W_{\hat{H}%
}^{(l)}(i,j)|^{2}C_{n}^{(l)}(j),  \label{25}
\end{eqnarray}%
where $N$ is the number of sources and $M$ the number of channels. The error $\Delta C_{\hat{H}}^{(l)}(i)$ refers to the frequency at which the mixing matrix has all its elements equal to 1. Eq.~
(\ref{25}) highlights how the different terms affect the estimation
error: the first term accounts for the contamination due to the other
sources, while the second one accounts for the effect of the
instrumental noise. Note that in the first term of eq.~(\ref{25}) there is the product of the reconstruction matrix  $\mathbf{W}_{\hat{H}}^{(l)}$, depending on the estimated mixing matrix $\mathbf{\hat H}$, with the true mixing matrix $\mathbf{H}$. As we will see, this term is minimum when $\mathbf{\hat H} = \mathbf{H}$ and increases as the estimated mixing matrix differs form the true one. From eq.~(\ref{25}), we can approximately evaluate the error on the spectrum of any component,
provided that the source and noise spectra are known, without
explicitly reconstructing that component.

Tegmark \& Efstathiou (1996) and Bouchet \&
Gispert (1999) performed an accurate analysis of eqs.~(\ref{23}) and
(\ref{25}) when the reconstruction matrix is a Wiener filter:
\begin{equation}
\mathbf{W}_{H}^{(l)} =\mathbf{C_{s}}^{(l)}\mathbf{H}^{T}[\mathbf{HC_{s}}^{(l)}
\mathbf{H}^{T} +\mathbf{C_{n}}^{(l)}]^{-1},  \label{26}
\end{equation}
where the mixing matrix is assumed as perfectly known, that is,
$\mathbf{\hat{H}=H}$.  They found that, although the Wiener filter
reduces the error due to instrumental noise, the contamination error
due to the other sources is always present, even if the actual matrix
$\mathbf{H}$ is known.  Moreover, the spectra estimated from the
sources, as given by eq.  (\ref{recon}), are biased.  The following
equation can be easily derived:
\begin{equation}
\mathbf{C_{\hat{s}}}^{(l)}=\frac{1}{2l+1}\sum_{m=-l}^{l}\langle \mathbf{\hat{%
s}}_{lm}\mathbf{\hat{s}}_{lm}^{\dagger }\rangle =\mathbf{W}_{H}^{(l)}\mathbf{%
HC_{s}}^{(l)}  \label{27}.
\end{equation}
Bouchet \& Gispert (1999) proposed to use a quality factor
derived from the quantity $\mathbf{W}_{H}^{(l)}\mathbf{H}$ in order to
correct the estimated CMB power spectrum.

If we reconstruct the sources by the following pseudo-inverse filter:
\begin{equation}
\mathbf{W}_{H}=\left( \mathbf{H}^{T}\mathbf{H}\right) ^{-1}\mathbf{H}^{T}
\label{28},
\end{equation}
we have $\mathbf{W}_{H}\mathbf{H=I}$.  Then, eq.~(\ref{23}) tells us
that the estimation error is only due to instrumental noise.
Moreover, the mean of the estimated power spectrum is:
\begin{equation}
\mathbf{C_{\hat{s}}}^{(l)}=\mathbf{C_{s}}^{(l)}+\mathbf{W}_{H}\mathbf{C}%
_{n}^{(l)}\left( \mathbf{W}_{H}\right) ^{\dagger }  \label{29}.
\end{equation}
It is apparent that this estimate is unbiased, but for large
multipoles the noise term becomes dominant on the reconstructed source
spectrum.  When the noise spectrum is known, as in the case of the
{\sc Planck} experiment, eq.~(\ref{29}) can be used to correct the
estimated spectra.

To evaluate the uncertainties on the estimated CMB
spectrum when the mixing matrix is not known exactly, we used a Monte Carlo approach: we generated 100 mixing matrices $\mathbf{\hat{H}}$ drawn from the probability distribution for the spectral indices described below. For each matrix we computed the $\Delta C_{\hat{H}}^{(l)}(cmb)$ from eq.~(\ref{25}).
For the LF set, these errors have been evaluated at the reference frequency of 100 GHz using
four sources (CMB, synchrotron, dust and free-free) and five channels
(30, 44, 70, 100, 143 GHz).  For the HF set we used two sources (CMB
and dust) and three channels (217, 353, 545 GHz), with the
reference frequency at 217 GHz.  For each channel set, we used the
all-sky power spectra of the reference source templates as
the input spectra $C_{s}^{(l)}(i)$. The $%
C_{n}^{(l)}(i)$ for each channel have been calculated as: 
\begin{equation}
C_{n}^{(l)}(i)=4\pi \sigma _{i}^{2}/N_{i},  \label{30}
\end{equation}%
where $N_{i}$ is the number of pixels, and the rms pixel noise values $\sigma
_{i}$ have been taken from Table \ref{tab:planck}.

 As we have been working with
sky patches, we should perform Monte Carlo simulations patch by patch,
using the appropriate basis, and then combine the information over the
whole sky.  However, since we only want to get an indicative amplitude
for $\Delta C_{\hat{H}}^{(l)}(cmb)$, we only performed our trials on a
single set of all-sky simulations, decomposed into
spherical harmonics. On the other hand, since an adequate strategy to
combine the results obtained on patches to achieve a coordinated
all-sky component separation has yet to be developed, we will
confine ourselves to the multipole range constrained by the patch
size.

For each iteration of our Monte Carlo simulations, we generated the mixing matrix from synchrotron and
dust spectral indices that are Gaussian-distributed 
around their true values  $\overline{\beta}_s$ and $\overline{\beta}_d$. In particular, we used standard deviations
of 0.1 and 0.05, respectively, for the distributions of synchrotron
and dust indices in the LF channels, and a standard deviation of 0.03
for the dust index in the HF channels.  These values are upper limits
to the ones shown in Fig.~\ref{fig:errvslat}. We then exploited eq.~(\ref{25})
 to compute the error on the CMB power spectrum for both the Wiener filter and the pseudo-inverse filter. 
From our Monte Carlo simulations, we  computed the average value of $\Delta C_{\hat{H}}^{(l)}(cmb)$, which is the total error on the CMB power spectrum estimation. We also estimated the standard deviation 
$\sigma _{\Delta C}$ of the random variable:
\begin{equation}
\Delta C_{\hat{H}}^{(l)}(cmb)-\Delta C_{H}^{(l)}(cmb),  \label{31}
\end{equation}%
where $\Delta C_{H}^{(l)}(cmb)$ is obtained for $\mathbf{\hat H} = \mathbf{H}$, and assumed it as the error on the CMB spectrum due to the
estimation of the mixing matrix. 

\begin{figure}
\begin{center}
\includegraphics[width=6cm, angle=90.]{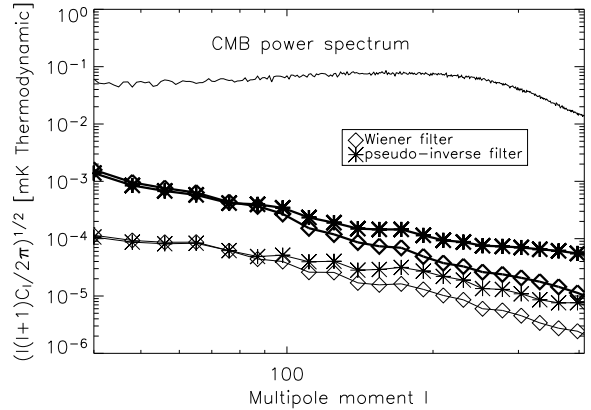}
\caption{Estimated errors on the CMB power spectrum at 100 GHz for
the LF set, for Wiener filter and pseudo-inverse filter. For each filter, the thicker (upper) line shows the total error, the lighter (lower) one is the contribution due to errors in the mixing matrix estimation.} \label{fig:cont_lf}
\end{center}
\end{figure}

\begin{figure}
\begin{center}
\includegraphics[width=6cm, angle=90.]{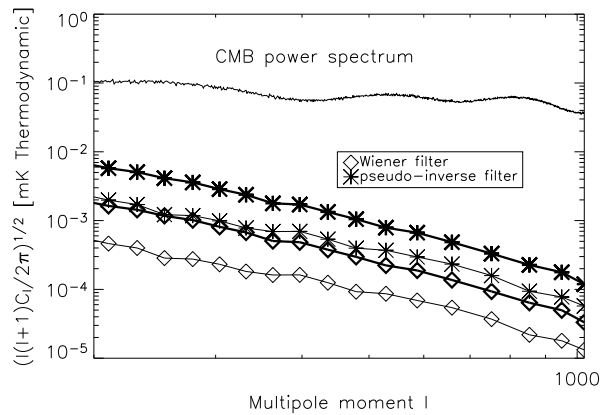}
\caption{Same as in Fig. \ref{fig:cont_lf}, but for the HF set, at the reference frequency of 217 GHz.} \label{fig:cont_hf}
\end{center}
\end{figure}

The results are shown in Figs. \ref{fig:cont_lf} and \ref{fig:cont_hf}. In Fig. \ref{fig:cont_lf} we show the errors for both the Wiener filter and the pseudo-inverse filter from the LF channel data, as functions of the multipole $l$, together with the original CMB
power spectrum. 
Figure \ref{fig:cont_hf} displays the corresponding
results for the HF channels. 
All the errors are computed without correction for the noise contribution; cosmic variance is not included.
Note that the Wiener filters are more accurate, but the differences with the pseudo-inverse solutions decrease once corrections for noise are applied. This is particularly true for the HF set, where the noise level is higher.
The total error on
the CMB power spectrum  $\Delta C_{\hat{H}}^{(l)}(cmb)$ turns out to be of the
order of $1\%$ for both the channels sets, and decreases from lower to higher multipoles.
In both cases the contribution of the errors in the estimation of
the mixing matrix is subdominant.

\section{Conclusions}

The tests described here demonstrate that the Correlated Component
Analysis (CCA) method (Bedini et al. 2005), applied to simulated data
with {\sc Planck} specifications, is a promising tool to estimate the 
mixing matrix parameterizing the frequency scaling of different 
astrophysical signals in cosmic microwave background (CMB) observations. 
For the low frequency {\sc Planck} channels, from 30 to
143 GHz, and with the angular resolution of $33'$, the most
accurate estimates are obtained in the latitude range
$[-30^\circ,+30^\circ]$, where the foregrounds we aim to recover
are relatively strong.

However, the strong concentration of intense free-free emission in
a narrow strip (a few degrees wide) around the Galactic equator
introduces a large gradient of the global spectral index of
Galactic radio emission (synchrotron plus free-free) that prevents
an accurate recovery of the synchrotron spectral index with the
low resolution of the LF channels. The problem is cured cutting
out the strip at $|b|< 3^\circ$. Having done that, both the
synchrotron and the dust spectral index are recovered with a mean
absolute error of 0.05. 
At higher latitudes, the mean error on the
synchrotron index increases to 0.08, while that on the dust index
is unchanged. 

The situation is even better with the high frequency {\sc Planck}
channels, from 217 to 545 GHz, thanks to the better spatial
resolution ($5'$) and to the fact that the only relevant
foreground, at least on intermediate angular scales, is Galactic
dust.

We have shown that, exploiting non-optimized component separation techniques, such as Wiener
filter and pseudo-inverse filter, such errors allow us to estimate the CMB power spectrum  with an uncertainty of the order of $1\%$ on the angular scales constrained by our patch size.

We conclude that the CCA can be a promising independent way to probe
the spectral behaviour of the main foregrounds affecting the CMB. This 
second order statistical approach may allow us to increase our 
knowledge on foregrounds and to perform component separation with
traditional non-blind methods with a minimum number of priors.
\section*{ACKNOWLEDGMENTS}
We gratefully acknowledge useful comments and suggestions from the
anonymous referee. Work supported in part by MIUR and ASI.


\begin{thebibliography}{}
\bibitem[]{} Aarts E., Korst J., 1989, \emph{Simulated Annealing and
Boltzmann Machines}, New~York, Wiley
\bibitem[]{} Amari S., Chichocki A., 1998, proc. IEEE 86, 2026
\bibitem[\protect\citeauthoryear{Baccigalupi et
al.}{2000}]{2000MNRAS.318..769B} Baccigalupi C. et al., 2000, MNRAS, 318, 769
\bibitem[\protect\citeauthoryear{Baccigalupi et
al.}{2004}]{2004MNRAS.354...55B} Baccigalupi C., Perrotta F., de
Zotti G., Smoot G.~F., Burigana C., Maino D., Bedini L., Salerno
E., 2004, MNRAS, 354, 55
\bibitem[\protect\citeauthoryear{Barreiro et
al.}{2004}]{2004MNRAS.351..515B} Barreiro R.~B., Hobson M.~P.,
Banday A.~J., Lasenby A.~N., Stolyarov V., Vielva P., G{\'o}rski
K.~M., 2004, MNRAS, 351, 515
\bibitem[]{} Bedini L.,Herranz D., Salerno E., Baccigalupi C., Kuruo\u{g}lu E.E., Tonazzini A., 2005, EURASIP Journal on Applied Signal Processing, 15, 2400

\bibitem[]{} Belouchrani A., Abed-Meraim K., Cardoso J.-F., Moulines E., 
1997, IEEE Trans. on signal Processing, 45, 434
%

\bibitem[\protect\citeauthoryear{Bennett et
al.}{2003}]{2003ApJS..148....1B} Bennett C.~L. et al., 2003a,
ApJS, 148, 1
\bibitem[\protect\citeauthoryear{Bennett et
al.}{2003}]{2003ApJS..148...97B} Bennett C.~L. et al., 2003b,
ApJS, 148, 97
\bibitem[]{} Bouchet F.R., Prunet S., Sethi S.K., 1999, MNRAS, 302,663
\bibitem[\protect\citeauthoryear{de Zotti et
al.}{1999}]{1999AIPC..476..204D} de Zotti G., Toffolatti L.,
Arg{\"u}eso F., Davies R.D., Mazzotta P., Partridge R.B., Smoot
G.F., Vittorio N., 1999, AIPC, 476, 204
\bibitem[]{} Dickinson C., Davies R.D., \& Davis R.J., 2003, MNRAS, 341 369

\bibitem[]{} Eriksen H.K., Dickinson C., Lawrence C.R., Baccigalupi
C., Banday A.J., G\'orski K.M., Hansen F.K., Lilje P.B., Pierpaoli E.,
Seiffert M.D., Smith K.M. \& Vanderlinde K., 2006, ApJ, 641, 665
%
\bibitem[]{} Finkbeiner D.P., Davies M., Schelegel D.J., 1999, ApJ 524, 867
\bibitem[]{} Finkbeiner D.P., 2003 ApJS, 146, 407
\bibitem[]{} Giardino G. et al., 2002, A\&A, 387, 82
\bibitem[]{} G\'orski K.M., Hivon E., Banday A.J., Wandelt B.D., Hansen F.K., 
Reinecke M., Bartelmann M., 2005, ApJ 622, 759 

\bibitem[]{} Haffner L.M., Reynolds R.J., Tufte S.L., 1999, ApJ, 523, 233
\bibitem[]{} Haslam C. G. T. et al., 1982, A\&A S 47,1
\bibitem[]{} Hivon E., G\'orski K.M., Netterfield C.B., Crill B.P., Prunet S., \& Hansen F., 2002, ApJ, 567,2
\bibitem[]{} Hyv\"arinen A. 1999, IEEE Signal Processing Lett. 6, 145
\bibitem[]{} Hobson M.P., Jones A.W., Lasenby A.N., Bouchet F., 1998, MNRAS, 300,1
\bibitem[\protect\citeauthoryear{Lamarre et
al.}{2003}]{2003NewAR..47.1017L} Lamarre J.~M. et al., 2003,
NewAR, 47, 1017
\bibitem[\protect\citeauthoryear{Maino et al.}{2002}]{2002MNRAS.334...53M}
Maino D. et al., 2002, MNRAS, 334, 53
\bibitem[\protect\citeauthoryear{Maino et al.}{2003}]{2003MNRAS.344..544M}
Maino D., Banday A.~J., Baccigalupi C., Perrotta F., G{\'o}rski
K.~M., 2003, MNRAS, 344, 544
\bibitem[\protect\citeauthoryear{Maisinger, Hobson, \&
Lasenby}{2004}]{2004MNRAS.347..339M} Maisinger K., Hobson M.~P.,
Lasenby A.~N., 2004, MNRAS, 347, 339

\bibitem[\protect\citeauthoryear{Mandolesi, Morgante, \&
Villa}{2003}]{2003SPIE.4850..722M} Mandolesi N., Morgante G.,
Villa F., 2003, SPIE, 4850, 722


\bibitem[\protect\citeauthoryear{Patanchon et
al.}{2005}]{2005MNRAS.364.1185P} Patanchon G., Cardoso J.-F.,
Delabrouille J., Vielva P., 2005, MNRAS, 364, 1185

\bibitem[]{} Schlegel D.J.,Finkbeiner D.P., Davies M., 1998, ApJ 500, 525

\bibitem[]{} Spergel D.N. et al., 2006, submitted to ApJS (astro-ph/0603449) 

\bibitem[]{} Stivoli S., Baccigalupi C., Maino D., Stompor R., 2006, 
submitted to MNRAS (astro-ph/0505381) 

\bibitem[\protect\citeauthoryear{Stolyarov et
al.}{2005}]{2005MNRAS.357..145S} Stolyarov V., Hobson M.~P.,
Lasenby A.~N., Barreiro R.~B., 2005, MNRAS, 357, 145

\bibitem[\protect\citeauthoryear{Tauber}{2004}]{2004AdSpR..34..491T} Tauber
J.A., 2004, AdSpR, 34, 491
\bibitem[]{} Tegmark M., 1996, MNRAS 280,299
\bibitem[]{} Tegmark M., Efstathiou G., 1996, MNRAS 281, 1297
\end{thebibliography}
\end{document}